# The Appell hypergeometric expansions of the solutions of the general Heun equation

A.M. Ishkhanyan

Institute for Physical Research, NAS of Armenia, 0203 Ashtarak, Armenia

Starting from the equation obeyed by the derivative, we construct several expansions of the solutions of the general Heun equation in terms of the Appell generalized hypergeometric functions of two variables of the fist kind. Several cases when the expansions reduce to ones written in terms of simpler mathematical functions such as the incomplete Beta function or the Gauss hypergeometric function are identified. The conditions for deriving finite-sum solutions via termination of the series are discussed. In general, the coefficients of the expansions obey four-term recurrence relations; however, there exist certain sets of the parameters for which the recurrence relations involve only two terms, though not successive. The coefficients of the expansions are then explicitly calculated and the general solution of the Heun equation is constructed in terms of the Gauss hypergeometric functions.

## Introduction

The general Heun equation [1], an equation that is currently widely applied in mathematics, physics, and engineering (see, e.g., [2-4] and references therein), is the most general ordinary linear second-order Fuchsian differential equation [5] having four regular singular points [2,3]. It directly generalizes the Gauss hypergeometric equation, which is the most general equation having three such points. Though usually in mathematics a generalization from three to four is not very significant, however, the case with the Heun equation seems to be a notable exception. The additional singularity introduces several qualitative complications as compared with the hypergeometric case [2,3]. For instance, the solutions are no more expressed in terms of definite or contour integrals involving simpler mathematical functions. Such integrals could be useful, e.g., in developing asymptotic expansions with respect to a variable parameter. Another major point is that the power-series expansions are no more governed by familiar two-term recurrence relations between the successive coefficients of the expansion [2,3]. As a result, the coefficients of the series are no more calculated explicitly and complications arise in studying of their convergence as well as in studying the connection problems for different expansions [2,3,6].

An alternative to the power series are the expansions in terms of more advanced mathematical functions. This is a general approach applicable to many differential equations including the confluent versions of the Heun equation. Such expansions have been suggested for the Heun class of equations and applied to different physical problems by many authors [7-20]. The used expansion functions include the Gauss hypergeometric function [7-12], the Kummer and Tricomi confluent hypergeometric functions [2,3,13-15], the functions of the Bessel class [15,16], Coulomb wave [17] and incomplete Beta [18-20] functions, etc. A few expansions in terms of the generalized hypergeometric functions, e.g., the Goursat function $_2F_2$ [21] and the Appell hypergeometric function of two variables of the first kind [22], are also known [20,23]. A useful by-product of these expansions are the finite-sum solutions derived by means of termination of the series. The most part of the exact solutions of the Heun equations known at the present time are derived namely in this way.

In choosing the expansion functions, the common approach applied to the general Heun equation in early papers was, based upon the immediate intuition, to use the Gauss hypergeometric functions with parameters matching the Heun equation as closely as possible. For this reason, functions having correct behavior at two singular points have been applied [7-11]. It has been shown, however, that functions which have correct behavior only at one singular point may also provide expansions [12,20] useful for applications in studying certain physical problems, e.g., the quantum two-state models (see [24,25]).

Such expansions came out in studying the properties of the derivatives of the solutions of the Heun equations [20,23,26-28]. It has been noticed that in some specific cases the equation obeyed by the derivative reduces to a Heun equation with altered parameters [20,23]. This is a useful observation since it allows one to construct several new expansions of the solutions of the Heun equations starting from the equations for their derivatives. Among these are the above-mentioned expansions in terms of the Goursat and Appell generalized hypergeometric functions, expansions in terms of combinations of the Beta-functions and the Gauss hypergeometric functions of a form not applied before [12,20]. However, these expansions apply to rather specific cases since in general the derivatives of the solutions of the Heun equations do not belong to the Heun class of functions [20,23,26].

In the present paper we go beyond and construct, starting from the equation obeyed by the derivative of the solution of the general Heun equation, several new expansions in terms of the Appell generalized hypergeometric functions of two variables of the first kind. The expansions are in general applicable for arbitrary sets of the involved parameters. The expansion coefficients obey four-term recurrence relations which in some cases reduce to ones involving three or two terms. We discuss the conditions for deriving finite-sum solutions by means of termination of the series, as well as identify several cases when the expansions reduce to ones written in terms of simpler mathematical functions such as the Beta or the hypergeometric function. In the cases when the recurrence relations involve only two terms, the coefficients of the expansions are explicitly calculated and the general solution of the Heun equation is constructed in terms of the Gauss hypergeometric functions.

**Expansions**

The general Heun equation is written as [1-3]

$$\frac{d^2u}{dz^2}+\left(\frac{\gamma}{z}+\frac{\delta}{z-1}+\frac{\varepsilon}{z-a}\right)\frac{du}{dz}+\frac{\alpha\beta z-q}{z(z-1)(z-a)}u=0, \qquad (1)$$

where the parameters satisfy the Fuchsian relation $1+\alpha+\beta=\gamma+\delta+\varepsilon$. This equation has four regular singular points located at $z=0,1,a$ and $\infty$. Its solution is denoted as $u=H(a,q;\alpha,\beta,\gamma,\delta;z)$ assuming that the value of $\varepsilon$ is determined from the Fuchsian relation. Note that below we assume that this notation refers to a solution of Eq. (1) defined up to an arbitrary constant multiplier, so that one can construct a solution normalized to the unity at the origin by dividing the written solution by $u(0)$ if this is finite and not zero.

It is readily verified by direct substitution that the function

$$v=z^{\gamma}(z-1)^{\delta}(z-a)^{\varepsilon}\frac{du}{dz} \qquad (2)$$

obeys the following second order equation:

$$\frac{d^2v}{dz^2}+\left(\frac{1-\gamma}{z}+\frac{1-\delta}{z-1}+\frac{1-\varepsilon}{z-a}-\frac{\alpha\beta}{\alpha\beta z-q}\right)\frac{dv}{dz}+\frac{\alpha\beta z-q}{z(z-1)(z-a)}v=0, \qquad (3)$$

which is a Fuchsian differential equation in general having five regular singular points - an additional singularity located at the point $z=q/(\alpha\beta)$ is manifested by the term $\alpha\beta/(\alpha\beta z-q)$. The singular points and the associated exponents are presented by the following Riemann $P$-symbol:

$$P\begin{pmatrix} 0 & 1 & a & q/(\alpha\beta) & \infty & \\ 0 & 0 & 0 & 0 & -\alpha & z \\ \gamma & \delta & \varepsilon & 2 & -\beta & \end{pmatrix}. \qquad (4)$$

It is immediately seen that in four particular cases, namely, if $q=0$, $q=\alpha\beta$, $q=a\alpha\beta$ and $\alpha\beta=0$, the point $q/(\alpha\beta)$ concedes with one of already existing singular points. Hence, in these four cases the number of singularities remains four, besides, the singularities remain regular so that in these cases Eq. (3) presents a general Heun equation with altered parameters as compared with Eq. (1). As it was already noted above, this property allows one to construct several expansions of the solutions of the Heun equation in terms of generalized hypergeometric functions, combinations of Beta-functions, and Gauss hypergeometric functions having correct behavior only in one singular point [20].

Here, extending the approach of [20,23], we present several expansions of the solutions of the general Heun equation in terms of the Appell generalized hypergeometric functions of two variables of the fist kind [22]. The expansions apply to different sets of the involved parameters, in general, arbitrary, not restricted to the above mentioned four particular cases. The idea is as follows. Consider a power-series expansion of a solution of Eq. (3) in the neighborhood of a point $z_0$ of complex plane, ordinary or singular, finite or infinite. For instance, let $z_0$ be a finite point and consider the following expansion:

$$v = (z-z_0)^\mu \sum_{n=0}^{+\infty} a_n (z-z_0)^n. \tag{5}$$

Substituting this series into Eq. (2) and integrating term by term, we arrive at the expansion

$$u = C_0 + \sum_{n=0}^{+\infty} a_n \left( \int z^{-\gamma}(z-1)^{-\delta}(z-a)^{-\varepsilon}(z-z_0)^{\mu+n} dz \right), \tag{6}$$

where $C_0$ is a constant. In many cases the integrals involved in this sum are expressed in terms of the Appell hypergeometric functions of two variables of the first kind defined by the double hypergeometric series [21,22]

$$F_1(\tilde{a};b_1,b_2;\tilde{c};x,y) = \sum_{m,n=0}^{+\infty} \frac{(\tilde{a})_{m+n}(b_1)_m(b_2)_n}{(\tilde{c})_{m+n}} \frac{x^m}{m!} \frac{y^n}{n!}, \tag{7}$$

where $(\tilde{a})_n$ is the Pochhammer symbol denoting the rising factorial. This series is absolutely convergent if $|x|,|y|<1$. If $\text{Re}(\tilde{c}) > \text{Re}(\tilde{a}) > 0$, the Appell function is also expressed by an Euler-type integral [21,22]:

$$F_1(\tilde{a};b_1,b_2;\tilde{c};x,y) = \frac{\Gamma(c)}{\Gamma(\tilde{a})\Gamma(c-\tilde{a})} \int_0^1 t^{\tilde{a}-1}(1-t)^{\tilde{c}-\tilde{a}-1}(1-xt)^{-b_1}(1-yt)^{-b_2} dt. \tag{8}$$

Thus, by applying Eqs. (5)-(6), several expansions of the solutions to the general Heun equation in terms of the Appell function (7) are constructed. The immediate four are the ones obtained by choosing $z_0$ as a singular point of the Heun equation (3), that is if $z_0 = 0, 1, a$ and $\infty$. Indeed, let, for instance, $z_0 = 0$, so that the expansion (5) presents a Frobenius solution of Eq. (3) in the neighborhood of its singular point $z = 0$:

$$v = z^\mu \sum_{n=0}^{+\infty} a_n^{(1)} z^n, \quad \mu = 0, \gamma. \tag{9}$$

In this case the expansion (6) reads

$$u = C_0 + \sum_{n=0}^{+\infty} a_n^{(1)} \left( \int z^{n+\mu-\gamma} (z-1)^{-\delta} (z-a)^{-\varepsilon} dz \right), \tag{10}$$

from where we get the series

$$u = C_0 + \sum_{n=0}^{+\infty} a_n^{(1)} u_n \tag{11}$$

with the following expansion functions (unless otherwise stated, we suppose $|z| \leq 1 \leq |a|$):

$$u_n = \frac{(-1)^{-\delta}}{(-a)^\varepsilon} \frac{z^{n+\gamma_0}}{n+\gamma_0} F_1\left( \gamma_0 + n; \delta, \varepsilon; 1 + \gamma_0 + n; z, \frac{z}{a} \right), \quad \mu = 0, \gamma, \tag{12}$$

where $\gamma_0 = 1 - \gamma + \mu$.

Similarly, choosing $z_0 = 1$ we get

$$u_n = \frac{(-1)^{-\delta+n+\mu}}{(-a)^\varepsilon} \frac{z^{1-\gamma}}{1-\gamma} F_1\left( 1-\gamma;\ \delta - \mu - n, \varepsilon;\ 2-\gamma;\ z, \frac{z}{a} \right), \quad \mu = 0, \delta. \tag{13}$$

The expansion corresponding to the choice $z_0 = a$ reads

$$u_n = \frac{(-1)^{-\delta}}{(-a)^{\varepsilon-n-\mu}} \frac{z^{1-\gamma}}{1-\gamma} F_1\left( 1-\gamma;\ \delta, \varepsilon - \mu - n;\ 2-\gamma;\ z, \frac{z}{a} \right), \quad \mu = 0, \varepsilon, \tag{14}$$

Finally, for $z_0 = \infty$ we have (compare with Eq. (12))

$$u_n = \frac{(-1)^{-\delta}}{(-a)^\varepsilon} \frac{z^{-n+\gamma_0}}{-n+\gamma_0} F_1\left( \gamma_0 - n; \delta, \varepsilon; 1 + \gamma_0 - n; z, \frac{z}{a} \right), \quad \mu = -\alpha, -\beta, \tag{15}$$

with $\gamma_0 = 1 - \gamma - \mu$.

If $z_0 = q/\alpha\beta$ or $z_0$ is an ordinary point of Eq. (3), the integrals involved in the expansion (6) in general are not expressed in terms of the Appell function, however, in several cases they are. This occurs, e.g., if one of the parameters $\gamma, \delta, \varepsilon$ is zero. For instance, suppose $\varepsilon = 0$ and consider the non-logarithmic solution of Eq. (3) in the neighborhood of the singular point $z_0 = q/\alpha\beta$. This solution, which corresponds to the larger exponent $\mu = 2$, see Eq. (4), if expanded in powers of $z - z_0$, results in expansion functions

$$u_n = \frac{(-1)^{-\delta}}{(-z_0)^{-2-n}} \frac{z^{1-\gamma}}{1-\gamma} F_1\left(1-\gamma;\ \delta,\ -2-n;\ 2-\gamma;\ z, \frac{z}{z_0}\right). \tag{16}$$

More advanced is the case when one of the parameters $\gamma, \delta, \varepsilon$ is a negative integer. Then applying a binomial expansion of the corresponding term, i.e., $z^{-\gamma}$, $(z-1)^{-\delta}$, $(z-a)^{-\varepsilon}$, respectively, the expansion is written as a sum of a finite number of series in terms of the Appell functions. For instance, in the simplest case $\varepsilon = -1$, obviously, we have the expansion

$$u = C_0 + \sum_{n=0}^{+\infty} a_n^{(5)}\left(u_{n+1} + (z_0 - a)u_n\right), \tag{17}$$

where $u_n$ is the expansion function given by Eq. (16).

**Expansions in terms of simpler mathematical functions**

In many cases the derived expansions are written in terms of simpler mathematical functions. An immediate option is the case when the second or the third parameter of the Appell function is zero. In this case the Appell function is reduced to an ordinary Gauss hypergeometric function:

$$F_1(\tilde{a}; b_1, 0; \tilde{c}; x, y) = {}_2F_1(\tilde{a}; b_1; \tilde{c}; x), \quad F_1(\tilde{a}; 0, b_2; \tilde{c}; x, y) = {}_2F_1(\tilde{a}; b_2; \tilde{c}; y). \tag{18}$$

For instance, if say $\varepsilon = 0$, using Eq. (12) we get

$$H(a,q;\alpha,\beta,\gamma,\delta;z) = C_0 + (-1)^{-\delta} \sum_{n=0}^{+\infty} a_n^{(1)} \frac{z^{n+\gamma_0}}{n+\gamma_0} {}_2F_1(n+\gamma_0, \delta; 1+n+\gamma_0; z), \tag{19}$$

where $C_0 = (-1)^{-\delta} a\mu/q$, $\gamma_0 = 1-\gamma+\mu$, $\mu = 0, \gamma$. Actually, in this case the expansion functions are further reduced to the incomplete Beta functions:

$$H(\varepsilon = 0) = C_0 + (-1)^{-\delta} \sum_{n=0}^{+\infty} a_n^{(1)} B(1+n-\gamma+\mu, 1-\delta; z), \quad \mu = 0, \gamma \tag{20}$$

Eqs. (12)-(16) suggest several immediate expansions of this form for $\delta = 0$ and $\varepsilon = 0$. For instance, from Eq. (13) we get another expansion for $\varepsilon = 0$:

$$H(\varepsilon = 0) = C_0 + (-1)^{-\delta} \sum_{n=0}^{+\infty} a_n^{(2)} (-1)^{n+\mu} B(1-\gamma, 1+n-\delta+\mu; z), \quad \mu = 0, \delta, \tag{21}$$

and Eq. (15) provides a third expansion:

$$H(\varepsilon = 0) = C_0 + (-1)^{-\delta} \sum_{n=0}^{+\infty} a_n^{(4)} B(1-n-\gamma-\mu, 1-\delta; z), \quad \mu = -\alpha, -\beta. \tag{22}$$

Similarly, Eqs. (12), (14) and (15) lead to the following expansions for $\delta = 0$:

$$H(\delta = 0) = C_0 + \sum_{n=0}^{+\infty} a_n^{(1)} \frac{a^{\gamma_n}}{(-a)^\varepsilon} B\left(\gamma_n, 1-\varepsilon; \frac{z}{a}\right), \quad \gamma_n = 1+n-\gamma+\mu, \quad \mu = 0, \gamma. \tag{23}$$

$$H(\delta = 0) = C_0 - \sum_{n=0}^{+\infty} a_n^{(3)} \frac{(-a)^{\varepsilon_n}}{a^\gamma} B\left(1-\gamma, \varepsilon_n; \frac{z}{a}\right), \quad \varepsilon_n = 1+n-\varepsilon+\mu, \quad \mu = 0, \varepsilon. \tag{24}$$

$$H(\delta = 0) = C_0 + \sum_{n=0}^{+\infty} a_n^{(4)} \frac{a^{\gamma_n}}{(-a)^\varepsilon} B\left(\gamma_n, 1-\varepsilon; \frac{z}{a}\right), \quad \gamma_n = 1-n-\gamma-\mu, \quad \mu = -\alpha, -\beta. \tag{25}$$

Using Eq. (6), similar expansions can be straightforwardly constructed also for $\gamma = 0$. Here are two such series:

$$H(\gamma = 0) = C_0 - \sum_{n=0}^{+\infty} a_n^{(2)} \frac{(a-1)^{\delta_n}}{(1-a)^\varepsilon} B\left(1-\varepsilon, \delta_n; \frac{a-z}{a-1}\right), \quad \delta_n = 1+n-\delta+\mu, \quad \mu = 0, \delta. \tag{26}$$

$$H(\gamma = 0) = C_0 + \sum_{n=0}^{+\infty} a_n^{(3)} \frac{(1-a)^{\varepsilon_n}}{(a-1)^\delta} B\left(\varepsilon_n, 1-\delta; \frac{a-z}{a-1}\right), \quad \varepsilon_n = 1+n-\varepsilon+\mu, \quad \mu = 0, \varepsilon. \tag{27}$$

Expansions in terms of the incomplete Beta functions of above form have been previously suggested for the above mentioned sets of parameters when the derivative of the solution of the Heun equation is a solution of another general Heun equation [20]. Furthermore, it has been shown that expansions in terms of such incomplete Beta functions can be constructed for certain sets of parameters also in the case of the confluent Heun equation [18]. Such expansions can be suggested also for many other equations, in particular, Fuchsian equations having more singular points.

It should be noted, however, that if any of these series in terms of Beta functions is terminated from the right hand-side, thus resulting in closed form solutions involving a finite number of Beta functions, the resultant solution is always written in terms of elementary functions, in general, quasi-polynomials [18]. This is a general property of Beta-function expansions of the solutions of any $k$ th order ordinary differential equation (linear or nonlinear) which can be written in the form $u = F(z, u', u'', ...., u^{(k)})$, where $F$ is an arbitrary elementary function of its arguments [18].

A complementary remark concerning the Beta-function expansions presented above is the following. The Gauss hypergeometric functions in terms of which the involved Beta functions are written, i.e. the hypergeometric functions involved in expansion (19) have the following Riemann $P$-symbol representation:

$$P\begin{pmatrix} 0 & 1 & \infty \\ 0 & 0 & n+\gamma_0 \\ -n-\gamma_0 & 1-\delta & \delta \end{pmatrix}. \tag{28}$$

It is seen that these hypergeometric functions have correct behavior only at one singular point, namely, at $z=1$. In this sense these functions differ from those used in earlier studies of the Gauss hypergeometric expansions of the solutions of the general Heun equation [7-8].

Another set of parameters for which the Appell function is simplified to an ordinary hypergeometric series is $b_1 = b_2$, $y = -x$. In this case the Appell function is reduced to a Clausen generalized hypergeometric function $_3F_2$ of the argument $z^2$:

$$F_1(\tilde{a}; b_1, b_1; \tilde{c}; x, -x) = {}_3F_2\left(\frac{1+\tilde{a}}{2}, \frac{\tilde{a}}{2}, b_1; \frac{1+\tilde{c}}{2}, \frac{\tilde{c}}{2}; x^2\right). \tag{29}$$

Interestingly, in our case, if we put $\delta = \varepsilon$ and $a = -1$ in Eq. (12) or Eq. (15), one of the upper parameters of the involved Clausen functions cancels a lower parameter, so that in these cases the expansion functions are further reduced to ordinary Gauss hypergeometric functions. For instance, the resultant expansion generated by $z_0 = 0$ reads

$$u = C_0 + (-1)^{-\delta} \sum_{n=0}^{+\infty} a_n^{(1)} \frac{z^{n+\gamma_0}}{n+\gamma_0} {}_2F_1\left(\frac{n+\gamma_0}{2}, \delta; 1+\frac{n+\gamma_0}{2}; z^2\right), \tag{30}$$

where $C_0 = (-1)^{-\delta} a\mu/q$, $\gamma_0 = 1 - \gamma + \mu$, and $\mu = 0, \gamma$. As it was in the case of expansion (19), here the hypergeometric functions are also further reduced to the incomplete Beta functions (this time, of the argument $z^2$):

$$u = C_0 + \frac{(-1)^{-\delta}}{2} \sum_{n=0}^{+\infty} a_n^{(1)} B\left(\frac{n+\gamma_0}{2}, 1-\delta; z^2\right). \tag{31}$$

Note that unlike expansion (19), the finite-sum solutions generated by termination of this series in general are not written in terms of elementary functions.

Several further cases of reductions of the Appell function to simpler mathematical functions are known [22]. Some cases can be deduced applying changes of the variable that leave unaltered the Heun equation (1) or the fundamental form of the integral in Eq. (6). For instance, the Heun equations with $a_1 = 2$ and $a = 1/2$ are readily transformed into a Heun equation with $a_1 = -1$ by applying $z_1 = z-1$ and $z_1 = 2z-1$, respectively.

In many cases the Appell functions reduce to combinations of simpler functions. For instance, if $a = -1$ and $\varepsilon = \delta - N$ with $N$ being a natural number, presenting $(z-a)^{-\varepsilon}$ as

$$(z-a)^{-\varepsilon} = (z+1)^{-\delta} \sum_{k=0}^{N} \binom{N}{k} z^k \tag{32}$$

we see that the corresponding expansion function (12) reduces to a linear combination of $N+1$ incomplete Beta functions of the argument $z^2$. For $N=1$ the result reads

$$u = C_0 + \frac{(-1)^{-\delta}}{2} \sum_{n=0}^{+\infty} a_n^{(1)} \left( B\left(\frac{n+\gamma_0}{2}, 1-\delta; z^2\right) + B\left(\frac{1+n+\gamma_0}{2}, 1-\delta; z^2\right) \right). \tag{33}$$

There are many other cases when the Appell function can be written in terms of other higher transcendental functions. Here is an example when the expansion is written in terms of the Hurwitz-Lerch transcendent [21]:

$$H(\delta = -\varepsilon = 1) = C_0 + \sum_{n=0}^{+\infty} a_n^{(1)} z^{n+\gamma_0} \left( \frac{1}{n+\gamma_0} + (a-1)\Phi(1, n+\gamma_0; z) \right), \tag{34}$$

where $C_0 = 2a\mu/q$, $\gamma_0 = 1-\gamma+\mu$, $\mu = 0, \gamma$. For the exponent $\mu = \gamma$ the second argument of the Hurwitz-Lerch function becomes integer: $n+\gamma_0 = 1+n$. As a result, the expansion functions are written as $u_n = P_n(z) - (a-1)\ln(1-z)$, where $P_n(z)$ is a polynomial.

**Recurrence relations for expansion coefficients**

In several cases the presented expansions involve a finite number of terms. To discuss when this is possible, consider, for instance, the Frobenius solution of Eq. (3) in the neighborhood of its singular point $z = 0$. It is readily shown that the coefficients $a_n^{(1)}$ of the expansion (9) are governed by the following *four-term* recurrence relation:

$$S_n a_n^{(1)} + R_{n-1} a_{n-1}^{(1)} + Q_{n-2} a_{n-2}^{(1)} + P_{n-3} a_{n-3}^{(1)} = 0, \tag{35}$$

where
$$S_n = -aq(n+\mu)(n-\gamma+\mu), \tag{36}$$

$$R_n = q^2 + (n+\mu)^2(q+aq+a\alpha\beta) - (n+\mu)(a\alpha\beta(1+\gamma) + q(\gamma+\varepsilon-1+a(\gamma+\delta-1))), \tag{37}$$

$$Q_n = -(n+\mu)^2(q+(1+a)\alpha\beta) + (n+\mu)(q(\gamma+\delta+\varepsilon-2) + \alpha\beta(\gamma+a\gamma+a\delta+\varepsilon)) - 2q\alpha\beta, \tag{38}$$

$$P_n = \alpha\beta(n-\alpha+\mu)(n-\beta+\mu) \tag{39}$$

with $\mu = 0, \gamma$. The expansion applies if $\gamma$ is not an integer. It is immediately seen that the recurrence relation becomes *three-term* if $q = 0$ or $\alpha\beta = 0$. Besides, in the exceptional case when $q = 0$, $a = -1$, $\varepsilon = \delta$, the relation involves only *two* terms since then $S_n$ and $Q_n$ both identically vanish for all $n$.

If $q \neq 0$, the series is left-hand side terminated at $n = 0$ if $S_0 = 0$, i.e., if $\mu = 0$ or $\mu = \gamma$. In this case for the limit $\rho = \lim_{n\to\infty}(a_n/a_{n-1})$ from Eq. (35) we get a cubic equation, the roots of which are 1, $1/a$ and $\alpha\beta/q$. According to Poincaré and Perron, $a_n/a_{n-1}$ generally tends to the larger root $\rho_{max}$ so that the series is convergent if $|z\rho_{max}|<1$. Hence, the convergence radius of the expansion (9) is $\min\{1,|a|,|q/(\alpha\beta)|\}$.

If the recurrence relation for $a_n^{(1)}$ is four-term, i.e., if both $q \neq 0$ and $\alpha\beta \neq 0$, the series (9) terminates from the right-hand side if three successive coefficients vanish. Let $a_N$ be the last non-zero coefficient and $a_{N+1} = a_{N+2} = a_{N+3} = 0$ for some $N = 1,2,\ldots$. We get that then should be $P_N = 0$ so that since $\alpha\beta \neq 0$, termination is possible if

$$\alpha = N + \mu \text{ or } \beta = N + \mu, \quad \mu = 0, \gamma. \tag{40}$$

The remaining two equations, $a_{N+1} = 0$ and $a_{N+2} = 0$, impose two more restrictions on the parameters. For $N = 1$ these conditions for $\mu = 0$ simplify to either $\{\delta = 0, q = \alpha\beta\}$ or $\{\varepsilon = 0, q = a\alpha\beta\}$, which present the simple cases when the Heun equation loses a singular point and so is reduced to the Gauss hypergeometric equation. Starting from $N = 2$, however, the results are not trivial.

If $q = 0$ and $\alpha\beta \neq 0$, the recurrence relation becomes three-term and the series is left-hand side terminated if $R_0 = 0$, i.e., if $\mu = 0$ or $\mu = 1 + \gamma$. In this case the point $z_0 = q/(\alpha\beta)$ coincides with the existing singularity $z = 0$ and the convergence radius becomes $\min\{1,|a|\}$.

As it was already mentioned above, in the exceptional case when $q = 0$ and additionally $a = -1$ and $\varepsilon = \delta$, both $S_n$ and $Q_n$ vanish for all $n$ so that the recurrence relation involves only two terms. In this case the coefficients $a_n^{(1)}$ are explicitly calculated:

$$a_n^{(1)} = \frac{1+(-1)^n}{2} c_{n/2}, \quad c_k = \frac{((\mu-\alpha)/2)_k ((\mu-\beta)/2)_k}{(1+\mu/2)_k ((1-\gamma+\mu)/2)_k}, \quad \mu = 0, 1+\gamma, \tag{41}$$

where $(\ldots)_k$ is the Pochhammer symbol, and the resultant expansion is eventually written in terms of the incomplete Beta functions of the argument $z^2$:

$$H(-1,0;\alpha,\beta,\gamma,\delta=\varepsilon;z) = \frac{\mu}{\alpha\beta} + \frac{1}{2}\sum_{k=0}^{+\infty} c_k B\left(\frac{1-\gamma+\mu}{2}+k, 1-\delta; z^2\right). \tag{42}$$

This expansion extends the one given by Eq. (31), which is applicable for $q \neq 0$, to the case $q = 0$. Note that the two expansions differ by the value of the constant $C_0$ which arises as a result of integration. We recall that we assumed that the notation $u = H(a,q;\alpha,\beta,\gamma,\delta;z)$ does not refer to a solution normalized at the origin to the unity but to a solution which is defined up to an arbitrary constant multiplier.

The cases when the coefficients of the expansions of the solutions of the Heun equation are governed by two-term recurrence relations are very rare. A particular set of the involved parameters when this occurs is discussed in Ref. [29]. In that exceptional case the values adopted by the Heun function at singular points $z=0$ and $z=1$ are explicitly calculated in terms of generalized hypergeometric functions. In the present case, if $\mu = 1+\gamma$, the solution (42) is restricted at the origin, the Beta functions all go to zero, so that we have

$$H(-1,0;\alpha,\beta,\gamma,\delta=\varepsilon;0) = \frac{1+\gamma}{\alpha\beta}. \tag{43}$$

For the value adopted at $z=1$ the result reads

$$H(-1,0;\alpha,\beta,\gamma,\delta=\varepsilon;1) = \frac{1+\gamma}{\alpha\beta} \frac{\Gamma((1+\gamma)/2)\Gamma(1-\delta)}{\Gamma((1+\gamma-\alpha)/2)\Gamma(1+\gamma-\beta)/2)}. \tag{44}$$

With these values, if divided by $(1+\gamma)/\alpha\beta$, Eq. (42) reproduces a known result, the first of the two fundamental solutions composing the general solution of the Heun equation, which is known for $a=-1$, $q=0$, $\varepsilon=\delta$ to be

$$H(-1,0;\alpha,\beta,\gamma,\delta=\varepsilon;z) = C_1 \cdot {}_2F_1\left(\frac{\alpha}{2},\frac{\beta}{2};\frac{1+\gamma}{2};z^2\right) + C_2 z^{1-\gamma} {}_2F_1\left(\delta-\frac{\alpha}{2},\delta-\frac{\beta}{2};\frac{3-\gamma}{2};z^2\right). \tag{45}$$

Correspondingly, if the second exponent $\mu = 0$ is applied, the expansion (42) reproduces the second term of this general solution.

Consider finally the recurrence relation for coefficients $a_n^{(5)}$ of expansion (5) of the solution of Eq. (3) in the neighborhood of the additional singular point $z_0 = q/(\alpha\beta)$, $\alpha\beta \neq 0$. This is again generally a four-term relation:

$$S_n a_n^{(5)} + R_{n-1} a_{n-1}^{(5)} + Q_{n-2} a_{n-2}^{(5)} + P_{n-3} a_{n-3}^{(5)} = 0 \tag{46}$$

with
$$S_n = z_0(z_0-1)(z_0-a)(n+\mu)(n-2+\mu), \tag{47}$$

$$R_n = (n+\mu)\left(z_0^2(3\mu_n-(\gamma+\delta+\varepsilon))+z_0(\gamma+\varepsilon-2\mu_n+a(\gamma+\delta-2\mu_n))+a(\mu_n-\gamma)\right), \tag{48}$$

$$Q_n = (n+\mu)\left(z_0(3(n+\mu)-2(\gamma+\delta+\varepsilon))+(\gamma-n-\mu)(1+a)+\varepsilon+a\delta\right), \tag{49}$$

$$P_n = (n-\alpha+\mu)(n-\beta+\mu), \tag{50}$$

where $\mu_n = n+\mu-1$. It is readily shown that the convergence radius of the expansion (5) with these coefficients equals to the distance from the point $z_0$ to the nearest other singular point of Eq. (3): $\min\{|z_0|,|z_0-1|,|z_0-a|\}$.

This recurrence relation involves only three successive terms only if $q=0$, $q=\alpha\beta$, $q=a\alpha\beta$. As already mentioned above, in these three cases $z_0$ coincides with one of already existing singular points so that in these cases the derivative of the solution of the general Heun equation is a solution of another Heun equation with altered parameters.

Except for these three specific cases, the series may left-hand side terminate if $\mu = 0$ or $\mu = 2$. Since the exponents differ by an integer, we have at least one consistent power-series expansion; the second solution will generally require a logarithmic term. The warranted power series is generated by the greater exponent $\mu = 2$; it is this exponent that leads to the expansion functions (16). It should be noted, however, that in certain cases the logarithmic solution also leads to expansions in terms of the Appell functions. For instance, this is the case if both $\delta$ and $\varepsilon$ are negative integers.

For termination of the series (5) from the right-hand side at some $N = 1, 2, \ldots$, should necessarily be $P_N = 0$, or, equivalently if $\alpha = N + 2$ or $\beta = N + 2$ for the exponent $\mu = 2$. The termination occurs if additionally $a_{N+1} = 0$ and $a_{N+2} = 0$. These conditions impose two more restrictions on the parameters.

In some cases the recurrence relation (46) may involve three, instead of four, terms, however, not successive. This occurs if one of the coefficients $R_n$ or $Q_n$ identically vanishes for all $n$. The coefficient $R_n$ vanishes if the parameters fulfill the following conditions:

$$3z_0^2 - 2(1+a)z_0 + a = 0, \quad (\gamma - \varepsilon)(z_0 - 1) + (\delta - \varepsilon)z_0 = 0, \tag{51}$$

and $Q_n$ vanishes if

$$z_0 = (1+a)/3, \quad (\gamma - 2\delta + \varepsilon) + a(\gamma + \delta - 2\varepsilon) = 0. \tag{52}$$

A notable feature of these two pairs of equations is that the first equation of each of the pair defines a specific choice for the point $z_0 = q/(\alpha\beta)$, thus fixing the singularities of Eq. (3), while the second equation implies a corresponding restriction on the exponents of the Heun equation. Interestingly, in the exceptional case when

$$a = (-1)^{\pm 1/3}, \quad q/(\alpha\beta) = (1+a)/3, \quad \gamma = \varepsilon = \delta, \tag{53}$$

both $R_n$ and $Q_n$ simultaneously vanish, so that in this specific case the recurrence relation (46) involves only two terms: $S_n a_n^{(5)} + P_{n-3} a_{n-3}^{(5)} = 0$. The coefficients $a_n^{(5)}$ are then explicitly calculated in terms of the Gamma functions (compare with Eq. (41)):

$$a_n^{(1)} = \frac{1 + a^{2n} + a^{-2n}}{3} c_{n/3}, \quad c_k = (3a)^{3k/2} \frac{\left((\mu-\alpha)/3\right)_k \left((\mu-\beta)/3\right)_k}{\left(1+\mu/3\right)_k \left((1+\mu)/3\right)_k}, \quad \mu = 0, 2, \tag{54}$$

Now, applying the expansion (16) and further doing the same steps as in deriving Eq. (45), we arrive at the following exact solution of the Heun equation for the parameters (53):

$$H(a,\alpha\beta(1+a)/3;\alpha,\beta,\gamma=\delta=\varepsilon;z) = C_1 \cdot {}_2F_1\left(\frac{\alpha}{3},\frac{\beta}{3};\frac{2}{3};-\frac{a^{3/2}(1+a-3z)^3}{3\sqrt{3}}\right) +$$
$$C_2(1+a-3z){}_2F_1\left(\frac{1+\alpha}{3},\frac{1+\beta}{3};\frac{4}{3};-\frac{a^{3/2}(1+a-3z)^3}{3\sqrt{3}}\right). \quad (55)$$

This reproduces a result by Maier [30]. It can be checked that many other similar results [30,31] can be derived using the above expansions in terms of the Appell hypergeometric functions. We will discuss these and other developments in a future publication.

**Acknowledgments**

This research has been conducted within the scope of the International Associated Laboratory (CNRS-France & SCS-Armenia) IRMAS. The research has received funding from the European Union Seventh Framework Programme (FP7/2007-2013) under grant agreement No. 205025 – IPERA.